\documentclass[conference]{IEEEtran}
\usepackage{cite}
\usepackage{amsmath,amssymb,amsfonts}
\usepackage{algorithm}
\usepackage{graphicx}
\usepackage{textcomp}
\usepackage{xcolor}
\usepackage{algpseudocode}
\usepackage{tabularx}
\usepackage{tikz}

\def\BibTeX{{\rm B\kern-.05em{\sc i\kern-.025em b}\kern-.08em
    T\kern-.1667em\lower.7ex\hbox{E}\kern-.125emX}}
\usepackage{soul,color}
\usepackage{multirow}

\usepackage{url}

\begin{document}

\title{Context-Aware Adaptive Prefetching for\\DASH Streaming over 5G Networks}

\author{\IEEEauthorblockN{Juncal Uriol\IEEEauthorrefmark{1}, Inhar Yeregui, \'Alvaro Gabilondo\IEEEauthorrefmark{1}, Roberto Viola}
\IEEEauthorblockA{\textit{Fundaci\'on Vicomtech} \\
\textit{Basque Research and Technology Alliance}\\
San Sebasti\'an, 20009 Spain\\
\{juriol, iyeregui, agabilondo, rviola\}@vicomtech.org}
\IEEEauthorrefmark{1}PhD Candidate at UPV/EHU
\and
\IEEEauthorblockN{Pablo Angueira, Jon Montalb\'an}
\IEEEauthorblockA{\textit{Department of Communications Engineering} \\
\textit{University of the Basque Country (UPV/EHU)}\\
Bilbao, 48013 Spain\\
\{pablo.angueira, jon.montalban\}@ehu.eus}

}

\maketitle

\IEEEoverridecommandlockouts
\IEEEpubid{\begin{minipage}{\textwidth}\ \\\\\\\\\\[12pt]\centering
J. Uriol et al., "Context-Aware Adaptive Prefetching for DASH Streaming over 5G Networks", 2023 IEEE International Symposium on Broadband Multimedia Systems and Broadcasting (BMSB), 2023, pp. 1-6, doi: 10.1109/BMSB58369.2023.10211275. \copyright 2023 IEEE. Personal use of this material is permitted. Permission from IEEE must be obtained for all other uses, in any current or future media, including reprinting/republishing this material for advertising or promotional purposes, creating new collective works, for resale or redistribution to servers or lists, or reuse of any copyrighted component of this work in other works.
\end{minipage}}

\begin{abstract}
The increasing consumption of video streams and the demand for higher-quality content drive the evolution of telecommunication networks and the development of new network accelerators to boost media delivery while optimizing network usage. Multi-access Edge Computing (MEC) enables the possibility to enforce media delivery by deploying caching instances at the network edge, close to the Radio Access Network (RAN). Thus, the content can be prefetched and served from the MEC host, reducing network traffic and increasing the Quality of Service (QoS) and the Quality of Experience (QoE). This paper proposes a novel mechanism to prefetch Dynamic Adaptive Streaming over HTTP (DASH) streams at the MEC, employing a Machine Learning (ML) classification model to select the media segments to prefetch. The model is trained with media session metrics to improve the forecasts with application layer information.
The proposal is tested with Mobile Network Operators (MNOs)' 5G MEC and RAN and compared with other strategies by assessing cache and player's performance metrics.
\end{abstract}


\section{Introduction}
\IEEEPARstart{N}{owadays}, video consumption is the cause of the largest amount of Internet traffic, while, at the same time, the demand for video content is also growing. These trends are pushed by the increasing number of users accessing media services and the explosion of video sensors and devices with improved video capabilities, such as high video resolution (Ultra-High-Definition or 4K) and high frame rate (HFR).
In this context, it is evident the need to enhance the network capabilities to target a certain level of Quality of Service (QoS) and Quality of Experience (QoE) required by each media service.

Dynamic Adaptive Streaming over HTTP (DASH) \cite{sodagar2011mpeg2} is the solution adopted to deliver video content while using the existing Content Delivery Networks (CDNs) without modifications. 
Nevertheless, an approach based only on CDN to serve DASH content presents some drawbacks.
First, the player strives to achieve the best individual quality, without knowledge of other connected players. This causes high network dynamics and unfairness in network utilization \cite{akhshabi2012}, which may lead to temporal interruptions and frequent changes in video representation. Ultimately, it may damage the QoE \cite{Seufert2014}.
Second, traffic generated by video consumption is redundant as the CDN has to stream a popular video as many times as the number of connected players. Thus, it affects the Content Provider (CP)'s Operational Expenditure (OPEX) \cite{silva2020}.

Multi-access Edge Computing (MEC) \cite{etsi2019}, including among 5G technologies, enables cloud capabilities at the network edge to enforce and boost the QoS/QoE of heterogeneous use cases \cite{etsigsmec002}, including video streaming ones \cite{Jiang2021}.
The Mobile Network Operator (MNO) or the CP can deploy specific network functions at the MEC hosts to improve media services. The idea is to employ analytic models and algorithms to extract useful information about the media sessions and/or make predictions on future events or performance. Information and predictions are later exploited to design services implementing more intelligent mechanisms for content caching or video transcoding. In particular, the use of forecasts allows services to act proactively to avoid or, at least, minimize the effects of network underperformance or any predicted problems.

This paper provides a novel solution for prefetching media segments at the edge when delivering DASH streams over a 5G network. The solution is achieved by providing the following relevant contributions:

\begin{itemize}
    \item Creation of a DASH application-layer dataset to train four different Machine Learning (ML) classification models and to select the best one in terms of accuracy. These models forecast player's media segment requests.
    \item Integration of the selected ML model into a prefetching mechanism at the edge that employs the forecasts 
    to make decisions on the segment to prefetch.
    \item Integration of the
    prefetching mechanism into Mobile Network Operators (MNOs)' 5G RAN and MEC infrastructure.
    \item Validation and comparison of the proposed solution with other prefetching strategies
    by assessing QoS and QoE performance indicators.
\end{itemize}


The rest of the paper is structured as follows. Section \ref{sec:sotaTBC} reviews the related work in the domain of prefetching and content caching applied to media delivery. Section \ref{sec:solutionTBC} describes article's main contribution, detailing the system architecture and the prefetching mechanism to boost the media delivery. Section \ref{sec:implementation} presents the setup where the solution is implemented and tested, and compiles the validation of the ML classification model. Section \ref{sec:results} details the results obtained by comparing our proposal with other caching strategies. Finally, we assert our conclusions and future work in Section \ref{sec:conclusions}.

\section{Related Work}
\label{sec:sotaTBC}

CDN is the worldwide solution employed by CPs to improve their media services. CDN aims at selectively replicating and caching the content at different Points of Presence (PoPs) such that the users can quickly access it from nearby locations.
This solution presents some highly significant limitations, especially when low-latency requirements come into play, as their location directly affects the overall latency.

In the literature, several edge caching solutions leveraging MEC architecture are proposed to overcome such limitations.
These solutions typically host a prefetching mechanism to cache the content before the actual request from the video player. In some cases, they can also be empowered with media segment and content popularity analysis \cite{Ge2016, chen2020}. These solutions also may decrease CP's OPEX, as they reduce redundant traffic from the CDN.

In \cite{viola2018}, a MEC proxy features local edge caching to reduce network traffic. It identifies the video player's requests and prefetches one media segment in advance at all the available representations.
A more complete analysis is presented in \cite{viola2022assessment}, where different caching strategies at the MEC node are compared. The authors found that a caching solution, including a prefetching mechanism, allows video players to move to higher-quality representations, but it may not be optimal regarding network consumption. When all the representations are prefetched at the MEC, some of them may never be requested by any connected video player.
In \cite{Tan2018}, the authors consider the Radio Network Information Service (RNIS), envisioned by ETSI MEC specifications \cite{etsigsmec012}, as an enabler to further improve content delivery. The radio information is exploited to select the representation to be prefetched at any time. Network state knowledge allows selectively prefetching the media segments at specific representations. To reduce the prefetched segments and, consequently, the amount of network traffic, the authors of \cite{kumar2018edge} propose to prefetch only the highest bitrate representation and transcode it to lower representation bitrates at the MEC host. As a drawback, it needs increased computing resources at the MEC in order to process the content.
Finally, a combination of prefetching and transcoding is considered in \cite{behravesh2022machine}. The cost of each operation (transcode and prefetch) is taken into account to find the best trade-off between them.


Our approach consists in empowering the prefetching operations at the MEC by including an ML classification model to forecast the next segment request. Therefore, the segment to be requested by the player is cached in advance, obtaining an intelligent and efficient caching system exploiting the metrics obtained from the media session.

\section{Context-aware adaptive prefetching}
\label{sec:solutionTBC}

\begin{figure}[ht]
\centering
    \includegraphics[width=0.5\textwidth,keepaspectratio]{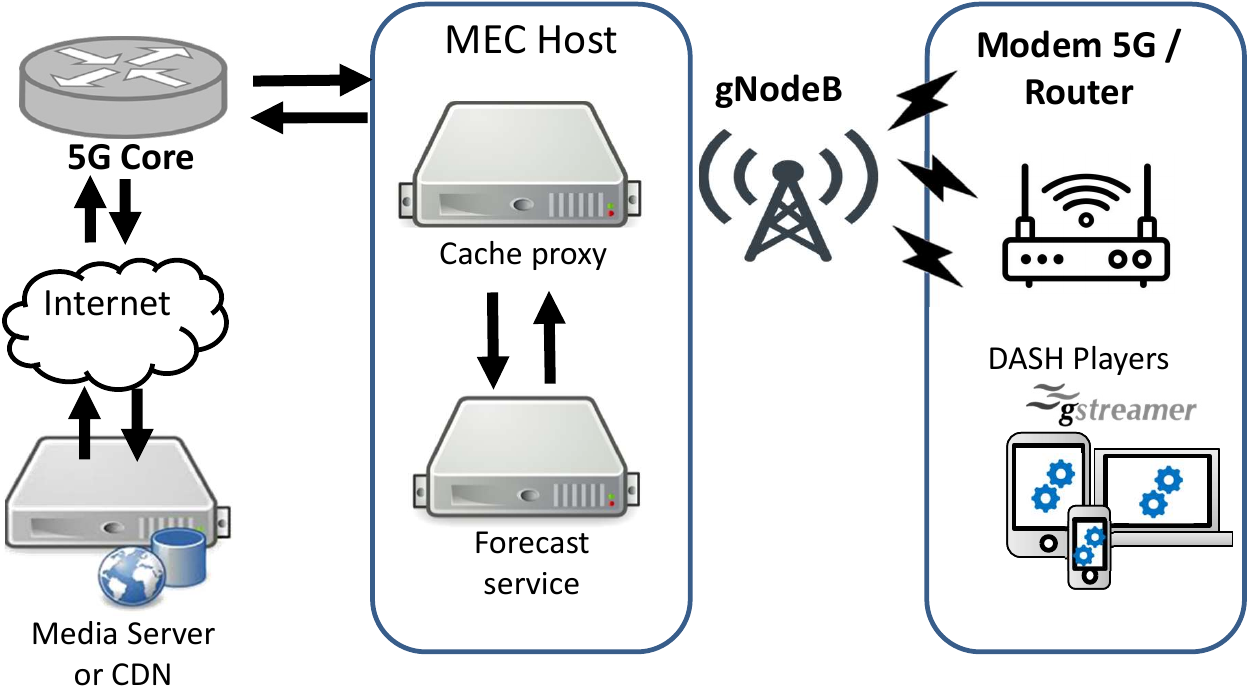}
\caption{System architecture for a MEC-enabled mobile network}
\label{fig:architecture}
\end{figure}

\subsection{System architecture} 
\label{sec:detailedcontribution}

Figure \ref{fig:architecture} shows the general architecture proposed to deliver DASH streams over 5G networks, implementing a forecast-powered prefetching at the MEC. The Media Presentation Description (MPD) and the media segments are stored and served by the Media Server or the CDN. The MPD is configured such that the \textit{BaseURL} addresses the Cache proxy deployed at the MEC node. Then, the DASH player analyzes the MPD and requests the media segments at the proxy, which retrieves them from the remote server (Media Server or CDN) before serving them.


In a simple approach where there is no caching of the media segments, the proxy downloads a segment only when it is actually requested by the player. On the contrary, when it implements a prefetching mechanism, it monitors some metrics of the streaming session to extract valuable information to feed a prediction model at a Forecast service. Then, the outcome of the model is employed by the proxy to cache the content in advance.

\subsection{Prefetching mechanism} \label{sec:prefetch_algorithm}
\begin{figure}[htp!]
    \centering
    \includegraphics[width=0.5\textwidth]{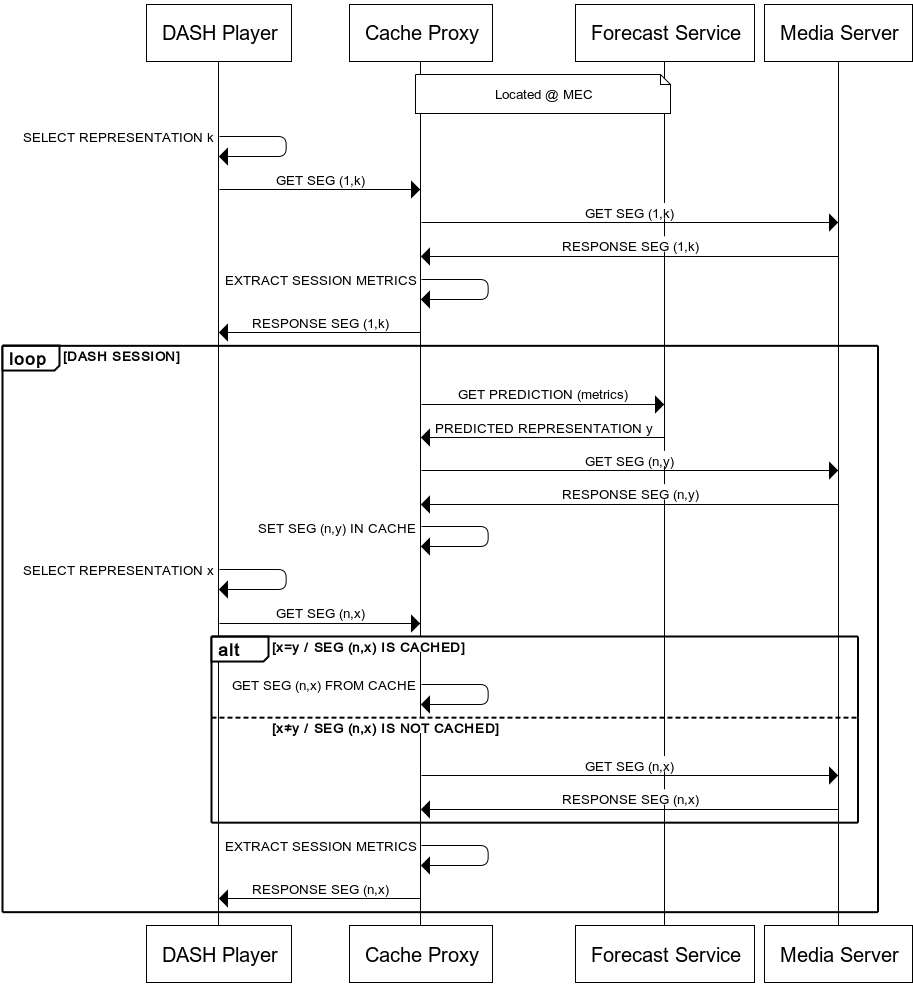}
    \caption{Prefetching Mechanism sequence diagram}
    \label{fig:seg_diagram}
\end{figure}

In order to improve the cache performance, the Forecast Service plays an important role. Its objective is to forecast the next segment representation such that it can be prefetched and cached before the player requests it. The prefetching sequence diagram is presented in Figure \ref{fig:seg_diagram}. When the first media segment is requested, the Cache Proxy download it from the media server, as there are no media session metrics yet to forecast the next segment representation. When serving it to the player, it extracts media session metrics that are used to feed the Forecast Service. 

For all the following segment requests, the Cache Proxy previously queries the Forecast service to predict the next segment representation.
Then, the segment is cached in the Proxy Cache at the predicted representation. Finally, when the player requests the media segment, if it is cached, it is sent to the player directly from the cache, but if it is not cached, it is requested to the media server and sent to the player.





\section{Implementation} \label{sec:implementation}

\subsection{Testbed setup}
\begin{figure}[htbp]
    \centering
    \includegraphics[width=0.5\textwidth,keepaspectratio]{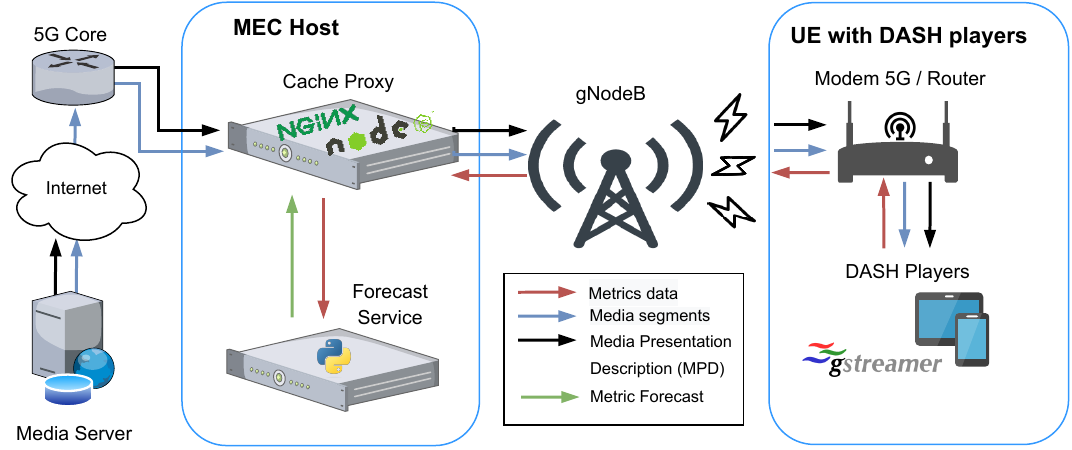}
    \caption{Low-level architecture scheme of the proposed solution}
    \label{fig:implementation}
\end{figure}

In order to test our proposal, we uses the testbed presented in Figure \ref{fig:implementation}.
The setup includes the following elements:
\begin{itemize}
    
    \item Media Server: a public server located at ATHENA Christian Doppler Laboratory, providing a multi-codec DASH dataset \cite{taraghi2022multi}.
    The selected video representations are shown in Table \ref{tab:representations} and the segment duration is set to 4 seconds.
    \item Cache Proxy: a HTTP proxy based on Node.js \cite{nodejs} and NGINX \cite{reese2008nginx} and located at the MEC host. It enables the prefetching and caching of media segments transferred between the Media Server and the players.
    \item Forecast service: a node at the MEC host having a Python implementation of an ML classification model to forecast the next media segment representation. The predictions are employed by the Cache Proxy to proactively prefetch the next segments.
    \item 5G Core, MEC Host and gNodeB: Euskaltel MNO's 5G Core network, virtualized MEC infrastructure, and Orange MNO's 5G base station.
    \item UE with DASH Player: multiple DASH players based on GStreamer multimedia framework \cite{gstreamer} that are executed in a 5G-connected UE. The 5G modem used in this implementation is a Telit FN980. 
\end{itemize}

\subsection{Machine Learning model evaluation}

\begin{table}[]
\caption{Set of DASH video representations}
\centering
\def\arraystretch{1.2}
\setlength\tabcolsep{2.5pt}
\label{tab:representations}
\begin{tabular}{|c|c|c|c|c|}
\hline
Index & Codec & Bitrate   & Resolution & Framerate \\ \hline
1     & HEVC  & 0.5 Mbps  & 640x360    & 24fps     \\
2     & HEVC  & 1.4 Mbps  & 1280x720   & 24fps     \\
3     & HEVC  & 5.5 Mbps  & 1920x1080  & 24fps     \\
4     & HEVC  & 11 Mbps   & 3840x2160  & 24fps     \\
5     & HEVC  & 20 Mbps   & 5120x2880  & 24fps     \\
6     & HEVC  & 27.5 Mbps & 7680x4320  & 24fps     \\ \hline
\end{tabular}
\end{table}

\begin{figure}[ht!]
    \centering
    \includegraphics[width=0.5\textwidth, keepaspectratio]{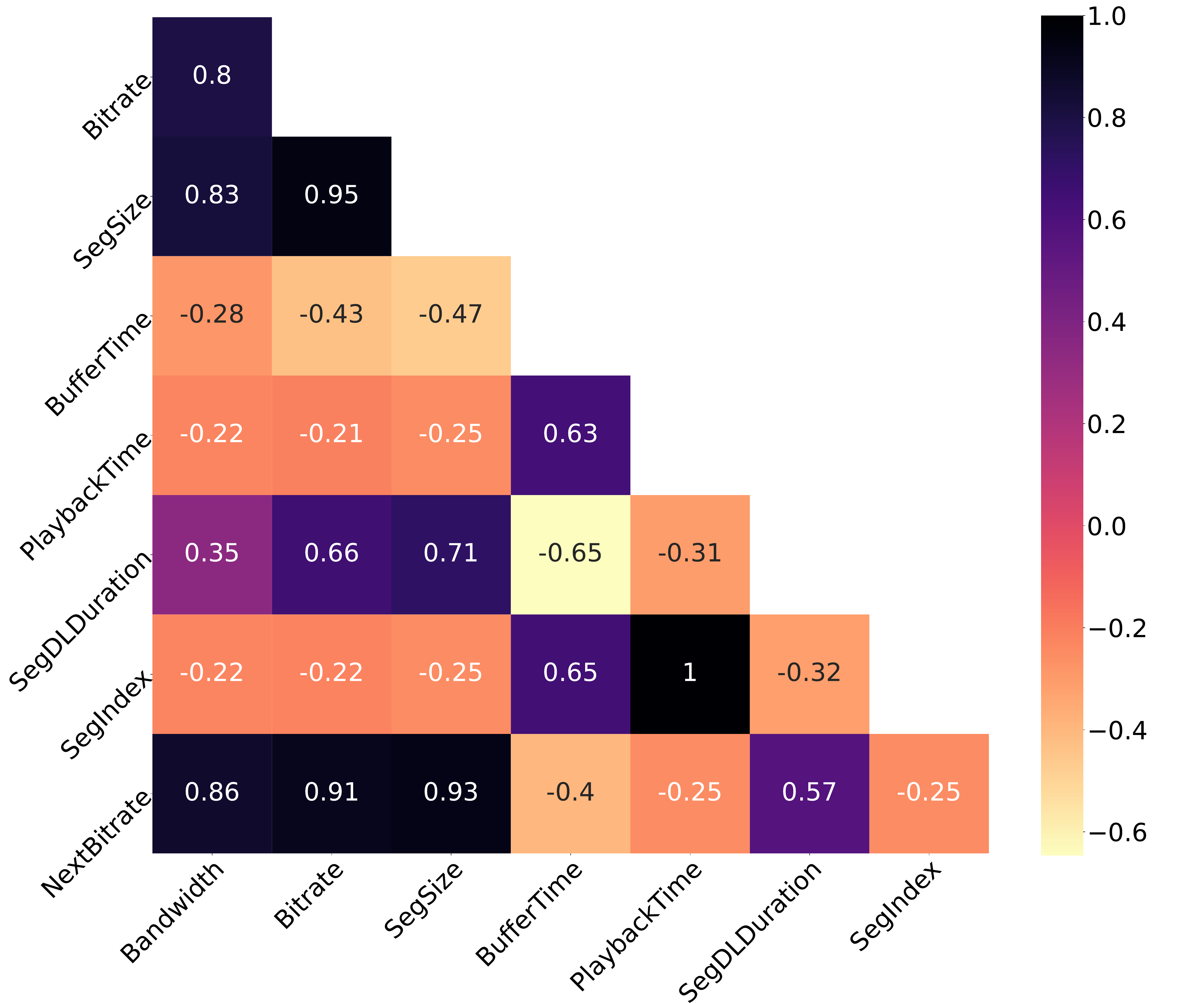}
    \caption{Correlation matrix of dataset metrics}
    \label{fig:correlation_heatmap_50}
\end{figure}

In this subsection, the design and validation of the ML classification model, located at the Forecast Service, are explained. Since the output of the ML model is known: 6 possible available bitrates in Mbps as indicated in Table \ref{tab:representations}, we have chosen an ML classification model for multi-class problems for predicting the next media segment representation, where the encoding bitrate can uniquely identify the representation.
A dataset has been generated by simulating 20 DASH players based on GStreamer multimedia framework over a 5G network and extracting media session metrics. The generated dataset has been used for the training and validation of the ML classification model.
The correlation heatmap in Figure \ref{fig:correlation_heatmap_50} is obtained with the generated dataset and presents the normalized correlation value between the media session metrics employed for the prediction, as indicated in the sequence diagram in Figure \ref{fig:seg_diagram}, and how they influence it.
The last row of the heatmap represents the output feature, i.e., the next segment representation bitrate, and its relation with the input metrics. It can be seen that the most influential metrics for predicting the next segment bitrate (\textit{NextBitrate}) are the network bandwidth (\textit{Bandwidth}), the current segment bitrate (\textit{Bitrate}) and the current segment size (\textit{SegSize}), all of them with a correlation value greater than 0.8.

We have compared four ML classification models: Random Forest Classifier (RF) \cite{belgiu2016random}, K-Neighbor Classifier (KN) \cite{manocha2007empirical}, Support Vector Machines (SVM) \cite{noble2006support} and Linear Discriminant Analysis (LDA) \cite{xanthopoulos2013linear}. RF gives good results in similar previous works \cite{behravesh2022machine}.  We explore KN because it is very easy to implement in multi-class problems, is robust to noisy data and is effective in large datasets \cite{imandoust2013application}. The SVM model works really well with a clear margin of separation, that is, the separation values of the output features, and we have studied it to see the model's effectiveness. Finally, we have studied the possibility of implementing the LDA because it works well with multi-class problems and is fast in terms of computation time \cite{starzacher2008evaluating}. 



\begin{table}[htp]
    \centering
    \caption{Parameters and accuracy of ML classification models}
    \begin{tabular}{|l|l|r|}
        \hline
        \textbf{ML model} & \textbf{Parameters} & \textbf{Accuracy (\%)}\\
        \hline
        \multirow{3}{5em}{RF} & Estimators: 100 & \multirow{3}{5em}{78.1}\\
        & Max Depth: None &\\
        & Min samples per leaf: 2 &\\
        \hline
        \multirow{2}{5em}{KN} & Neighbors: 5 & \multirow{2}{5em}{75.0}\\
        & Weights: None &\\
        \hline
        \multirow{2}{5em}{SVM} & C: 1.0 & \multirow{2}{5em}{73.1}\\
        & Kernel: Radial Basis Function &\\
        \hline
        \multirow{2}{5em}{LDA} & Solver: Single Value Decomposition & \multirow{2}{5em}{69.0}\\
        & Shrinkage: None &\\
        \hline
    \end{tabular}
    \label{tab:model_parameters}
\end{table}

The definition of the parameters used to test the four ML classification models is resumed in Table \ref{tab:model_parameters}. These parameters are the ones given by default by the ML library. Each model has been trained and validated with the same generated dataset mentioned above in order to calculate the accuracy of each one.

Table \ref{tab:model_parameters} shows the accuracy of these four ML classification models. In order to select the best ML classification model for the prediction of the next segment bitrate, a comparison between ML models' accuracy is presented.
We have imposed an accuracy threshold of 75\% for the model validation. Even if all ML classification models are close to this accuracy threshold, two of them, specifically the SVM and LDA, do not reach it, so they have been immediately discarded. The other two models, the RF and KN, reach the accuracy threshold.
Since the RF gives the best result, we decide to implement it in the MEC for forecasting the next segment bitrate.



\begin{figure}[ht!]
    \centering
    \includegraphics[width=0.5\textwidth,keepaspectratio]{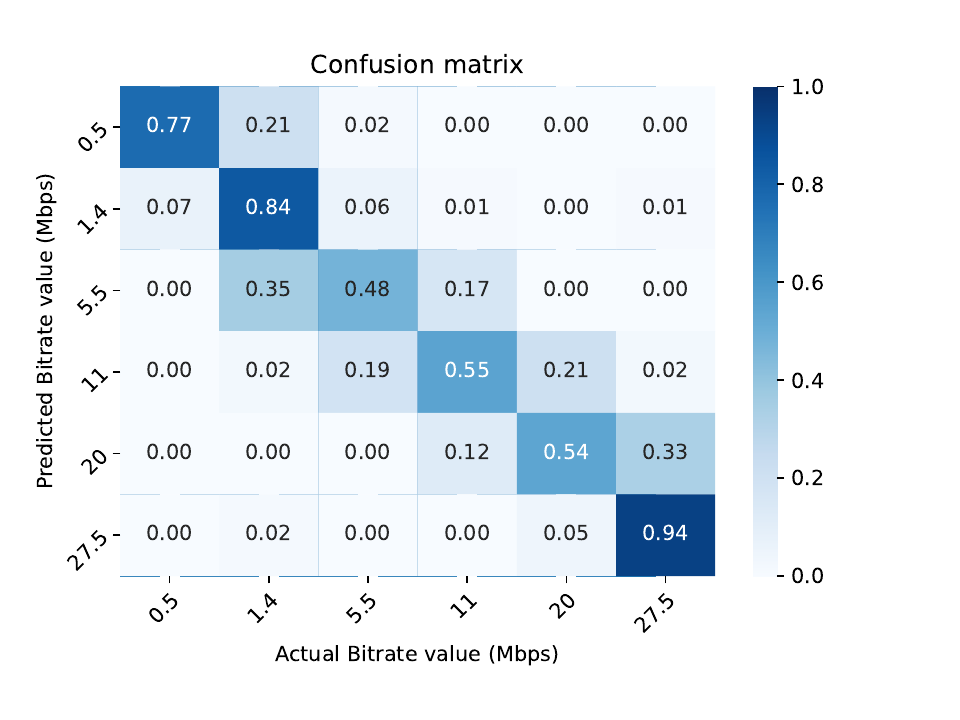}
    \caption{Ground truth normalized confusion matrix for RF classification model}
    \label{fig:confusion_matrix}
\end{figure}

Going into more detail in the chosen RF model, a confusion matrix is presented in Figure \ref{fig:confusion_matrix}, detailing the relationship between predicted and actual bitrates. The figure shows the probability that each representation bitrate is correctly predicted.

\section{Results}
\label{sec:results}

\begin{figure*}[htbp]
    \centering
    \includegraphics[width=1\textwidth, keepaspectratio]{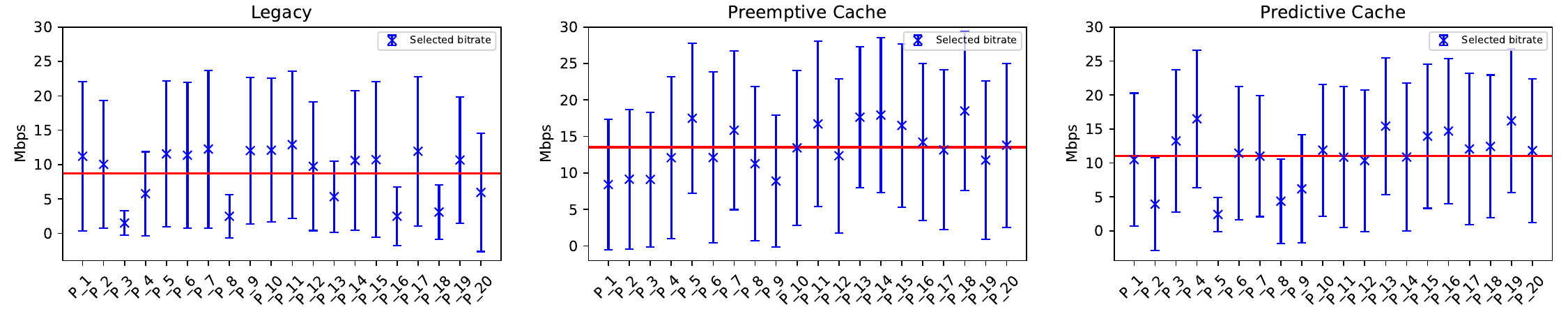}
    \caption{20 players sharing a radio link: average value and deviation of selected representation bitrate for different caching strategies.}
    \label{fig:datarate}
\end{figure*}


This section evaluates the proposed mechanism to prefetch DASH media segments at the MEC. In order to test the effects of the solution, three caching scenarios have been considered:

\begin{itemize}
    \item Legacy: the media segments are never cached. When a player requests a media segment, the proxy downloads it from the Media Server and serves it to the player.
    \item Preemptive Cache: the proxy prefetches all the representations of the next segment in advance. It means that the next player's request will be already cached.
    \item Predictive Cache: the proxy prefetches the next segment only at the predicted representation. The Forecast Service analyzes the media session metrics to predict the representation of the next player's request.
\end{itemize}

In both Preemptive Cache and Predictive Cache, the segments are prefetched approximately one segment duration (4 seconds) before the player's request and stored in cache during two segment duration (8 seconds) in order to ensure that the player has enough time to download it before being removed from the cache.
However, every time a player requests a cached segment, the segment is revalidated and remains in the cache for 8 more seconds after the player's request. This means that each segment is removed only when it is not requested by any of the players for 8 seconds.

Tests are carried out for each scenario by executing 20 players,
modelling their inter-arrival time through a modified version of the Poisson distribution \cite{consul1973generalization}. Their video playback is 322 seconds, i.e., the full video sequence length. 

\begin{table}[htp]
\caption{Cache proxy performance metrics.}
\centering
\def\arraystretch{1.2}
\setlength\tabcolsep{2.5pt}
\label{tab:cache}
\begin{tabular}{l|c|c|c|c|}
\cline{2-5}
& \textit{Hit$_{ratio}$} & \textit{Cached data} & \textit{Served data} & \textit{Data saved }\\ 
& & \textit{(GB)} & \textit{(GB)} & \textit{(\%)} \\ 
\hline
\multicolumn{1}{|l|}{\textbf{Legacy}}           & -         & -                 & 6.06  & -               \\ \hline
\multicolumn{1}{|l|}{\textbf{Preemptive Cache}} & 0.96      & 8.70                  & 8.39   & -3.65         \\ \hline
\multicolumn{1}{|l|}{\textbf{Predictive Cache}} & 0.73      & 4.56                 & 7.28  & 37.25         \\ \hline
\end{tabular}
\end{table}

Table \ref{tab:cache} shows the Cache Proxy performance metrics. It describes cache hit ratio (\textit{Hit$_{ratio}$}) and the information on data transferred over the Cache Proxy node, namely \textit{Cached data}, \textit{Served data} and \textit{Data saved}, for each considered scenario. Cached data represents the data transferred from the Media Server to the Cache Proxy in order to be cached. Served data represents the data transferred from the Cache to the players. In Legacy scenario, Served data will be data that the proxy obtains from the CDN at the moment that the players require it. Meanwhile, in Preemtive and Predictive Cache scenario, this data will already be cached in the proxy. Data saved expresses the relation percentage between both Cached and Served data.
The Legacy scenario does not cache any segments, so \textit{Hit$_{ratio}$} and \textit{Cached data} do not apply, while \textit{Served data} is 6.06 GB.
When employing a Preemptive cache, \textit{Hit$_{ratio}$} is 0.96, meaning that 96\% of the segment requests are already cached when served to the players. As backward, the cache is not actually reducing the traffic over the network, as \textit{Cached data} (8.70 GB) is 3.65\% bigger than \textit{Served data} (8.39 GB). Thus, \textit{Data saved} results in a negative value, as the traffic is increasing by using this strategy. This is reasonable as the Preemptive Cache is prefetching more segments than needed, as some cached segments are never requested, causing the usage of unnecessary network resources.
When employing the Predictive cache, \textit{Hit$_{ratio}$} is 0.73, meaning that its capability to serve segments from the cache is 24\% less than the Preemptive cache (0.96).
However, the Predictive cache is effectively saving 37.25\% of traffic data, as only 4.56 GB are cached in order to serve 7.28 GB to the players.
This means that the cached data is exploited in a more profitable way. We could conclude that the Predictive cache is the most effective caching strategy in terms of network usage.
Moreover, we could argue that Preemptive and Predictive strategies have higher transferred data than Legacy ones. This is not due to any loss of efficiency of using the cache, but due to the fact that the introduction of the cache allows the players to switch to higher-quality representations, requiring higher data transfer. This is evident when assessing QoS and QoE performance at the players.

\begin{table}[htp]
\caption{Player performance metrics}
\centering
\bgroup
\def\arraystretch{1.2}
\setlength\tabcolsep{2.5pt} 
\label{tab:player}
\begin{tabular}{l|c|c|c|c|c|c|}
\cline{2-7}
& \textit{R$_{avg}$} & \textit{S$_{n}$} & \textit{Stall$_{n}$} & \textit{Stall$_{avg}$} & \textit{QoE$_{avg}$} & \textit{QoE$_{dev}$} \\
& \textit{(Mbps)} & &  & \textit{(s)} & &\\
\hline
\multicolumn{1}{|l|}{\textbf{Legacy}}           & 8.69       & 37.00   & 1.60    & 6.00    &   4.05     &   0.27 \\
\hline
\multicolumn{1}{|l|}{\textbf{Preemptive Cache}}    & 13.51      & 36.75 & 1.05   & 4.47  & 4.38  &   0.20\\
\hline
\multicolumn{1}{|l|}{\textbf{Predictive Cache}} & 11.00      & 39.50  & 1.25   & 5.70     &   4.31  &   0.28 \\
\hline
\end{tabular}
\egroup
\end{table}

Table \ref{tab:player} describes players' performance metrics for each cache strategy. Concerning selected video representations, it is evident that the employment of a cache at the MEC provides the player with a higher network throughput, and therefore, they are able to download higher-quality representation segments. Figure \ref{fig:datarate} shows the average value and deviation of the bitrate of the downloaded segments for each player in each scenario. The red lines indicate the average value among all the players for each scenario, as reported in Table \ref{tab:player}. These average values show that, as expected, the Legacy strategy provides the lowest average bitrate (8.69 Mbps), while the Preemptive one gives the best one (13.51 Mbps).
As reasonable, the Predictive cache obtains an intermediate value of 11.00 Mbps.
The tendency of the values remains the same when considering the average number of stalls and their average duration. The Legacy strategy gives the worst values, while the Preemptive cache gives the best ones.
The average number of representation switches presents instead some differences. Legacy and Preemptive strategies show almost the same results (around 37 switches per player), while the Predictive one is 5\% worse (39.5 switches per player). Since the Predictive strategy is caching only a part of the overall segments, it results in a more unstable network from the player's point of view, causing issues in the player's adaptation algorithm \cite{viola2022assessment}.

Finally, information on selected representation, experienced stalls, and switches are inferred to obtain the QoE scores according to ITU-T P.1203 recommendation \cite{Robitza2018}. The average QoE for the Legacy strategy is 4.05, the lowest value among the three strategies. Preemptive cache QoE is 4.38, and Predictive cache QoE is 4.31. It means that players with Predictive cache have only 1.5\% less QoE score compared to Preemptive cache. When it comes to the deviation of the QoE score, the three values are similar. They are 0.27, 0.20, and 0.28 for Legacy, Preemptive and Predictive strategies, respectively. 

To sum up, the Predictive cache strategy results in a better solution to balance the trade-off between cache performance and player's QoE. Predictive cache enables the players to have a higher QoE compared to Legacy cache, and is similar to Preemptive cache. Moreover, the QoE improvement with the Predictive strategy comes at a much lower cost in terms of network data traffic compared to the Preemptive one.





\section{Conclusions and Future Work} \label{sec:conclusions}

This paper aims to present a novel prefetching mechanism by forecasting the next DASH media segment requested by the player. All the tests have been performed over MNOs' 5G network, where a MEC host is configured for forecasting and caching media segments thanks to media session information. The forecasts are performed by an ML classifier, chosen by assessing the accuracy of four different ones proposed in the literature.



The results show that the use of forecasts influences the Cache Hit Ratio and the Data Saved when caching the content at the MEC. By comparing three different caching strategies, we conclude that players tend to have a better QoS and QoE performance when a cache is employed. Moreover, a predictive cache results in a more efficient solution, as it allows to reduce network usage.




In the future, we plan to exploit new metrics to improve the predictions made by the forecast service. By considering physical wireless link and network layer information, further metrics may be collected to train a more accurate ML classification model.

\section*{Acknowledgment}


This research was supported by Red.es, Spain's 5G National Plan, under grant C012/12-SP for the 5G Euskadi project, and by Smart Networks and Services Joint Undertaking under the European Union’s Horizon Europe Research and Innovation programme, under Grant Agreement 101096838 for 6G-XR project.



\bibliographystyle{IEEEtran}
\bibliography{main.bib}

\begin{thebibliography}{10}
\providecommand{\url}[1]{#1}
\csname url@samestyle\endcsname
\providecommand{\newblock}{\relax}
\providecommand{\bibinfo}[2]{#2}
\providecommand{\BIBentrySTDinterwordspacing}{\spaceskip=0pt\relax}
\providecommand{\BIBentryALTinterwordstretchfactor}{4}
\providecommand{\BIBentryALTinterwordspacing}{\spaceskip=\fontdimen2\font plus
\BIBentryALTinterwordstretchfactor\fontdimen3\font minus
  \fontdimen4\font\relax}
\providecommand{\BIBforeignlanguage}[2]{{%
\expandafter\ifx\csname l@#1\endcsname\relax
\typeout{** WARNING: IEEEtran.bst: No hyphenation pattern has been}%
\typeout{** loaded for the language `#1'. Using the pattern for}%
\typeout{** the default language instead.}%
\else
\language=\csname l@#1\endcsname
\fi
#2}}
\providecommand{\BIBdecl}{\relax}
\BIBdecl

\bibitem{sodagar2011mpeg2}
I.~Sodagar, ``The mpeg-dash standard for multimedia streaming over the
  internet,'' \emph{IEEE multimedia}, vol.~18, no.~4, pp. 62--67, 2011.

\bibitem{akhshabi2012}
S.~Akhshabi, L.~Anantakrishnan, A.~C. Begen, and C.~Dovrolis, ``What happens
  when http adaptive streaming players compete for bandwidth?'' in
  \emph{Proceedings of the 22nd international workshop on Network and Operating
  System Support for Digital Audio and Video}, 2012, pp. 9--14.

\bibitem{Seufert2014}
M.~Seufert, S.~Egger, M.~Slanina, T.~Zinner, T.~Ho{\ss}feld, and P.~Tran-Gia,
  ``A survey on quality of experience of http adaptive streaming,'' \emph{IEEE
  Communications Surveys \& Tutorials}, vol.~17, no.~1, pp. 469--492, 2014.

\bibitem{silva2020}
S.~Da~Silva, J.~Bruneau-Queyreix, M.~Lacaud, D.~Negru, and
  L.~R{\'e}veill{\`e}re, ``Muslin: A qoe-aware cdn resources provisioning and
  advertising system for cost-efficient multisource live streaming,''
  \emph{International Journal of Network Management}, vol.~30, no.~3, p. e2081,
  2020.

\bibitem{etsi2019}
D.~Sabella, V.~Sukhomlinov, L.~Trang, S.~Kekki, P.~Paglierani, R.~Rossbach,
  X.~Li, Y.~Fang, D.~Druta, F.~Giust \emph{et~al.}, ``Developing software for
  multi-access edge computing,'' \emph{ETSI white paper}, vol.~20, pp. 1--38,
  2019.

\bibitem{etsigsmec002}
\BIBentryALTinterwordspacing
ETSI. (2018) Etsi gs mec 002: Multi-access edge computing (mec): Phase 2: Use
  cases and requirements. [Online]. Available:
  \url{https://www.etsi.org/deliver/etsi_gs/MEC/001_099/002/02.01.01_60/gs_MEC002v020101p.pdf}
\BIBentrySTDinterwordspacing

\bibitem{Jiang2021}
X.~Jiang, F.~R. Yu, T.~Song, and V.~C. Leung, ``A survey on multi-access edge
  computing applied to video streaming: Some research issues and challenges,''
  \emph{IEEE Communications Surveys \& Tutorials}, 2021.

\bibitem{Ge2016}
C.~Ge, N.~Wang, S.~Skillman, G.~Foster, and Y.~Cao, ``Qoe-driven dash video
  caching and adaptation at 5g mobile edge,'' in \emph{Proceedings of the 3rd
  ACM Conference on Information-Centric Networking}, 2016, pp. 237--242.

\bibitem{chen2020}
Y.~Chen, Y.~Liu, J.~Zhao, and Q.~Zhu, ``Mobile edge cache strategy based on
  neural collaborative filtering,'' \emph{IEEE Access}, vol.~8, pp.
  18\,475--18\,482, 2020.

\bibitem{viola2018}
R.~Viola, A.~Martin, M.~Zorrilla, and J.~Montalb{\'a}n, ``Mec proxy for
  efficient cache and reliable multi-cdn video distribution,'' in \emph{2018
  IEEE International Symposium on Broadband Multimedia Systems and Broadcasting
  (BMSB)}.\hskip 1em plus 0.5em minus 0.4em\relax IEEE, 2018, pp. 1--7.

\bibitem{viola2022assessment}
R.~Viola, D.~Amendola, Z.~Fern{\'a}ndez, {\'A}.~Gabilondo, M.~Zorrilla,
  P.~Angueira, M.~Casals, and J.~Montalb{\'a}n, ``Assessment of the effects of
  5g mec cache on dash adaptation algorithms,'' in \emph{2022 IEEE
  International Symposium on Broadband Multimedia Systems and Broadcasting
  (BMSB)}.\hskip 1em plus 0.5em minus 0.4em\relax IEEE, 2022, pp. 1--6.

\bibitem{Tan2018}
Y.~Tan, C.~Han, M.~Luo, X.~Zhou, and X.~Zhang, ``Radio network-aware edge
  caching for video delivery in mec-enabled cellular networks,'' in \emph{2018
  IEEE Wireless Communications and Networking Conference Workshops
  (WCNCW)}.\hskip 1em plus 0.5em minus 0.4em\relax IEEE, 2018, pp. 179--184.

\bibitem{etsigsmec012}
\BIBentryALTinterwordspacing
ETSI. (2017) Etsi gs mec 012: Mobile edge computing (mec); radio network
  information api. [Online]. Available:
  \url{https://www.etsi.org/deliver/etsi_gs/MEC/001_099/012/01.01.01_60/gs_MEC012v010101p.pdf}
\BIBentrySTDinterwordspacing

\bibitem{kumar2018edge}
S.~Kumar, D.~S. Vineeth \emph{et~al.}, ``Edge assisted dash video caching
  mechanism for multi-access edge computing,'' in \emph{2018 IEEE International
  Conference on Advanced Networks and Telecommunications Systems (ANTS)}.\hskip
  1em plus 0.5em minus 0.4em\relax IEEE, 2018, pp. 1--6.

\bibitem{behravesh2022machine}
R.~Behravesh, A.~Rao, D.~F. Perez-Ramirez, D.~Harutyunyan, R.~Riggio, and
  M.~Boman, ``Machine learning at the mobile edge: The case of dynamic adaptive
  streaming over http (dash),'' \emph{IEEE Transactions on Network and Service
  Management}, 2022.

\bibitem{taraghi2022multi}
B.~Taraghi, H.~Amirpour, and C.~Timmerer, ``Multi-codec ultra high definition
  8k mpeg-dash dataset,'' in \emph{Proceedings of the 13th ACM Multimedia
  Systems Conference}, 2022, pp. 216--220.

\bibitem{nodejs}
\BIBentryALTinterwordspacing
Node.js: asynchronous event driven javascript runtime. [Online]. Available:
  \url{https://nodejs.org/en/}
\BIBentrySTDinterwordspacing

\bibitem{reese2008nginx}
W.~Reese, ``Nginx: the high-performance web server and reverse proxy,''
  \emph{Linux Journal}, vol. 2008, no. 173, p.~2, 2008.

\bibitem{gstreamer}
\BIBentryALTinterwordspacing
Gstreamer: open source multimedia framework. [Online]. Available:
  \url{https://gstreamer.freedesktop.org/}
\BIBentrySTDinterwordspacing

\bibitem{belgiu2016random}
M.~Belgiu and L.~Dr{\u{a}}gu{\c{t}}, ``Random forest in remote sensing: A
  review of applications and future directions,'' \emph{ISPRS journal of
  photogrammetry and remote sensing}, vol. 114, pp. 24--31, 2016.

\bibitem{manocha2007empirical}
S.~Manocha and M.~A. Girolami, ``An empirical analysis of the probabilistic
  k-nearest neighbour classifier,'' \emph{Pattern Recognition Letters},
  vol.~28, no.~13, pp. 1818--1824, 2007.

\bibitem{noble2006support}
W.~S. Noble, ``What is a support vector machine?'' \emph{Nature biotechnology},
  vol.~24, no.~12, pp. 1565--1567, 2006.

\bibitem{xanthopoulos2013linear}
P.~Xanthopoulos, P.~M. Pardalos, and T.~B. Trafalis, ``Linear discriminant
  analysis,'' in \emph{Robust data mining}.\hskip 1em plus 0.5em minus
  0.4em\relax Springer, 2013, pp. 27--33.

\bibitem{imandoust2013application}
S.~B. Imandoust, M.~Bolandraftar \emph{et~al.}, ``Application of k-nearest
  neighbor (knn) approach for predicting economic events: Theoretical
  background,'' \emph{International journal of engineering research and
  applications}, vol.~3, no.~5, pp. 605--610, 2013.

\bibitem{starzacher2008evaluating}
A.~Starzacher and B.~Rinner, ``Evaluating knn, lda and qda classification for
  embedded online feature fusion,'' in \emph{2008 International Conference on
  Intelligent Sensors, Sensor Networks and Information Processing}.\hskip 1em
  plus 0.5em minus 0.4em\relax IEEE, 2008, pp. 85--90.

\bibitem{consul1973generalization}
P.~C. Consul and G.~C. Jain, ``A generalization of the poisson distribution,''
  \emph{Technometrics}, vol.~15, no.~4, pp. 791--799, 1973.

\bibitem{Robitza2018}
W.~Robitza, S.~Göring, A.~Raake, D.~Lindegren, G.~Heikkilä, J.~Gustafsson,
  P.~List, B.~Feiten, U.~Wüstenhagen, M.-N. Garcia, K.~Yamagishi, and
  S.~Broom, ``{HTTP Adaptive Streaming QoE Estimation with ITU-T Rec. P.1203
  – Open Databases and Software},'' in \emph{9th ACM Multimedia Systems
  Conference}, Amsterdam, 2018.

\end{thebibliography}

\end{document}